\documentclass[%
 reprint,
superscriptaddress,
nofootinbib,
nobibnotes,
nobibnotes,
 amsmath,amssymb,
 aps,
prb,
floatfix,
]{revtex4-2}
\usepackage{graphicx}
\usepackage{float}
\usepackage{epstopdf}
\usepackage{dcolumn}
\usepackage{bm}
\usepackage{gensymb}
\usepackage{upgreek}
\usepackage{hyperref}
\usepackage{tabularx}
\usepackage{placeins}
\usepackage{xcolor}
\usepackage{soul}
\preprint{APS/123-QED}
\usepackage{ulem}

\begin{document}

\title{Resilient $j$=3/2 superconductivity in topological semimetal YPtBi}

\author{Prathum Saraf}
\affiliation{Maryland Quantum Materials Center, Department of Physics, University of Maryland, College Park, Maryland 20742, USA}

\author{Nicholas A. Crombie}
\affiliation{Maryland Quantum Materials Center, Department of Physics, University of Maryland, College Park, Maryland 20742, USA}

\author{Rahul Sharma}
\affiliation{Maryland Quantum Materials Center, Department of Physics, University of Maryland, College Park, Maryland 20742, USA}

\author{Jared Z. Dans}
\affiliation{Maryland Quantum Materials Center, Department of Physics, University of Maryland, College Park, Maryland 20742, USA}

\author{Danila Sokratov}
\affiliation{Maryland Quantum Materials Center, Department of Physics, University of Maryland, College Park, Maryland 20742, USA}

\author{Carsyn L. Mueller}
\affiliation{Maryland Quantum Materials Center, Department of Physics, University of Maryland, College Park, Maryland 20742, USA}

\author{Ram Kumar}
\affiliation{Maryland Quantum Materials Center, Department of Physics, University of Maryland, College Park, Maryland 20742, USA}

\author{Hyunsoo Kim}
\affiliation{Maryland Quantum Materials Center, Department of Physics, University of Maryland, College Park, Maryland 20742, USA}
\affiliation{Department of Physics, Missouri University of Science and Technology, Rolla, Missouri 65409, USA}

\author{Connor Roncaioli}
\affiliation{Maryland Quantum Materials Center, Department of Physics, University of Maryland, College Park, Maryland 20742, USA}

\author{Winslow Weiss}
\affiliation{Maryland Quantum Materials Center, Department of Physics, University of Maryland, College Park, Maryland 20742, USA}

\author{David Graf}
\affiliation{National High Magnetic Field Laboratory, Tallahassee, FL 32310, USA}

\author{Chandra Shekhar}
\affiliation{Max Planck Institute for Chemical Physics of Solids, 01187 Dresden, Germany}

\author{Claudia Felser}
\affiliation{Max Planck Institute for Chemical Physics of Solids, 01187 Dresden, Germany}

\author{Johnpierre Paglione}
\email{paglione@umd.edu}
\affiliation{Maryland Quantum Materials Center, Department of Physics, University of Maryland, College Park, Maryland 20742, USA}
\affiliation{Canadian Institute for Advanced Research, Toronto, Ontario M5G 1Z8, Canada}

\begin{abstract}
Cooper pairing in most of the known fermionic superfluids occurs via  spin-1/2 quasiparticle interactions that lead to spin-singlet or spin-triplet pairing. In the topological semimetal YPtBi, strong spin-orbit coupling results in a band inversion between highly symmetric $s$- and $p$-like electronic bands and a degeneracy at the $\Gamma$ point that ensures the manifold of $j$=3/2 quasiparticle states thrive near the Fermi level, where superconducting pairing occurs.
Here we study the effects of magnetic and nonmagnetic disorder and carrier density on this exotic superconducting pairing state. By varying levels of disorder and carrier densities by nearly two and three orders of magnitude, respectively, we show that the superconducting critical temperature of YPtBi has a remarkable robustness, with little variation across this span. Our results suggest that superconductivity in YPtBi may reside in a regime where phase stiffness, rather than pair formation, governs the transition temperature. The insensitivity of Cooper pairing to dramatic changes in quasiparticle environment in a $j$=3/2 superconductor highlights a new form of protection realized in topological high-spin superconductors.
\end{abstract}

\date{\today}

\maketitle



The topological semimetal YPtBi is a prominent member of the half-Heusler family known to host inverted bands, strong spin-orbit coupling, and unconventional superconductivity~\cite{Chadov2010,Lin2010,Xiao2010}. Superconductivity below 1~K was first reported in YPtBi despite an exceptionally low carrier density of only $n \sim10^{18}$~cm$^{-3}$~\cite{ButchPRB,HKimSciAdv,HKimQO-PRS}, far below that expected for conventional electron-phonon pairing~\cite{MeinertPRL}. Similar to the case of dilute superconductor SrTiO$_3$, which has a similar carrier density  (albeit with much lower $T_c$) \cite{LinSrTiO3-PRX}, the survival of superconductivity in this limit is in contrast to basic expectations for a conventional superconductor where theory would predict vanishingly small transition temperatures.
However in YPtBi, strong spin-orbit coupling and the inversion of the $\Gamma_6$ and $\Gamma_8$ bands stabilize quasiparticles with dominant $j$=3/2 character near the Fermi level~\cite{BrydonPRL,HKimQO-PRS}, allowing a richer manifold of pairing channels, including singlet $(J=0)$, triplet $(J=1)$, quintet $(J=2)$ and septet $(J=3)$ states~\cite{HKimSciAdv,BrydonPRL,JYuPRB,SavaryPRB}. 
Together with evidence of nodal excitations from penetration depth measurements~\cite{HKimPRB,HKimSciAdv}, the possibility of unconventional topological superconductivity in this otherwise unassuming material 
has spurred great interest~\cite{RoyPRB,VenderbosPRX,JYuPRB,WangPRB,TimmPRB,SimPRB, 
Suzuki2016,Hirschberger2016,Manna2018,Roy2019,Roy2021,Kobayashi2022,JYuJAP,Dakyeong2023,Persky2025}.

A central unresolved question is how such a low-density $j$=3/2 superconducting state responds to perturbations that strongly modify the normal state electronic structure. In conventional superconductors, disorder and carrier density tuning primarily affect superconductivity through changes in the density of states or pair-breaking. Here, we investigate the response of YPtBi to disorder, hole doping, and magnetic substitution using electrical transport, Hall effect, and quantum oscillation measurements. Despite dramatic changes in scattering, conductivity, and Fermi surface properties, superconductivity remains remarkably robust. This resilience suggests that superconductivity in YPtBi is not governed solely by conventional amplitude-based pairing considerations, but instead may derive substantial stability from the phase stiffness and topology of the underlying superconducting state.

To systematically investigate the effects of disorder and doping on both the superconductivity and normal state electronic properties of YPtBi, we compare the response of electrical transport, quantum oscillations and Hall effect measurements in five types of samples: conventional YPtBi, ultra-high purity YPtBi, chemically disordered YPtBi, hole-doped (Y,Ca)PtBi and magnetically-substituted (Y,Nd)PtBi.
These representative cases induce a remarkably wide span of responses in elastic scattering rate (resistivity) and carrier density, while retaining the same character of $j$=3/2 Fermi surface, and the lack of sensitivity to these perturbations in the superconductivity of YPtBi points to an unprecedented nature of pairing in this system.


 \begin{figure*}[t]
 \includegraphics[width=0.7\linewidth]{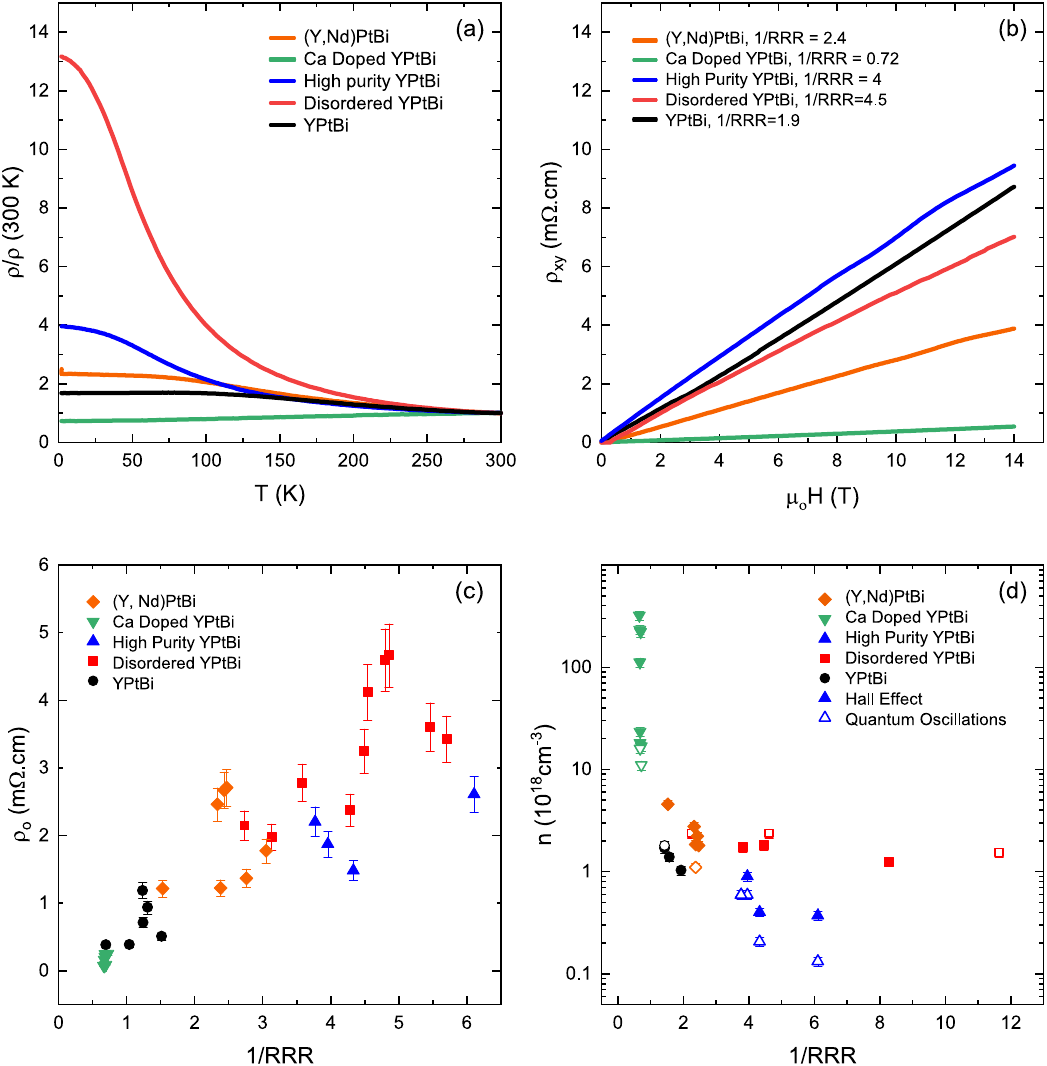}
    \hfill
    \caption{
    {\bf Variation of transport behavior of YPtBi with disorder, purity and charge doping.}
    (a) and (b) Normalized resistivity and Hall resistivity of the various types of YPtBi grown respectively. Panel (a) shows data for  Y$_{0.997}$Ca$_{0.003}$PtBi and Y$_{0.52}$Nd$_{0.48}$PtBi, while panel (b) show data for Y$_{0.991}$Ca$_{0.009}$PtBi and Y$_{0.80}$Nd$_{0.20}$PtBi. (c) Plot of 1/RRR vs residual resistivity indicating a change in the residual resistivity in the samples as opposed to room temperature resistivity. (d) Carrier density of various samples as a function of 1/RRR. Filled symbols are quantities measured with Hall and open symbols are measured with quantum oscillations. }
    \label{fig:Transport}
\end{figure*}

 \begin{figure*}[t]
    \centering
    \includegraphics[width=0.98\linewidth]{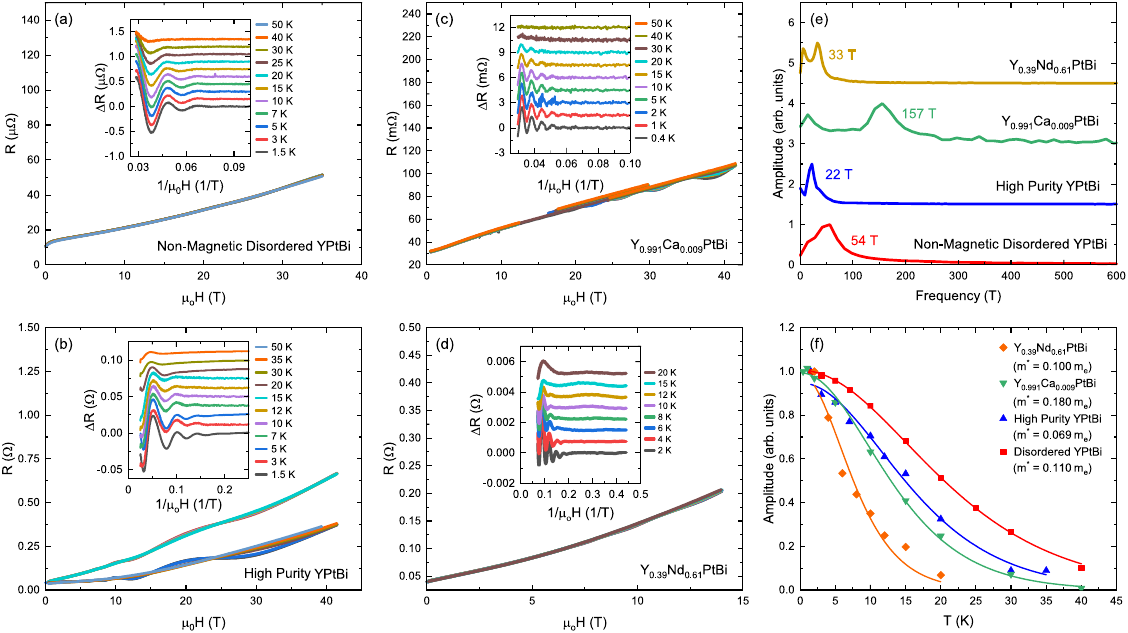}
    \caption{
    {\bf Quantum oscillations in chemically tuned YPtBi.}
    (a) - (d) Raw data of resistance vs magnetic field of disordered, high purity, Ca doped samples and (Y,Nd)PtBi respectively. Inset shows subtracted resistance curves offset to show the oscillations more clearly. (e) Fast Fourier transform of the oscillations showing changes in frequency of the four types of YPtBi. (f) Lifshitz-Kosevich fits (lines) of the temperature dependence of the amplitude of oscillations which is used to extract the effective mass.}
    \label{fig:QO}
\end{figure*}

\begin{figure*}[t]
    \centering
    \includegraphics[width=0.7\linewidth]{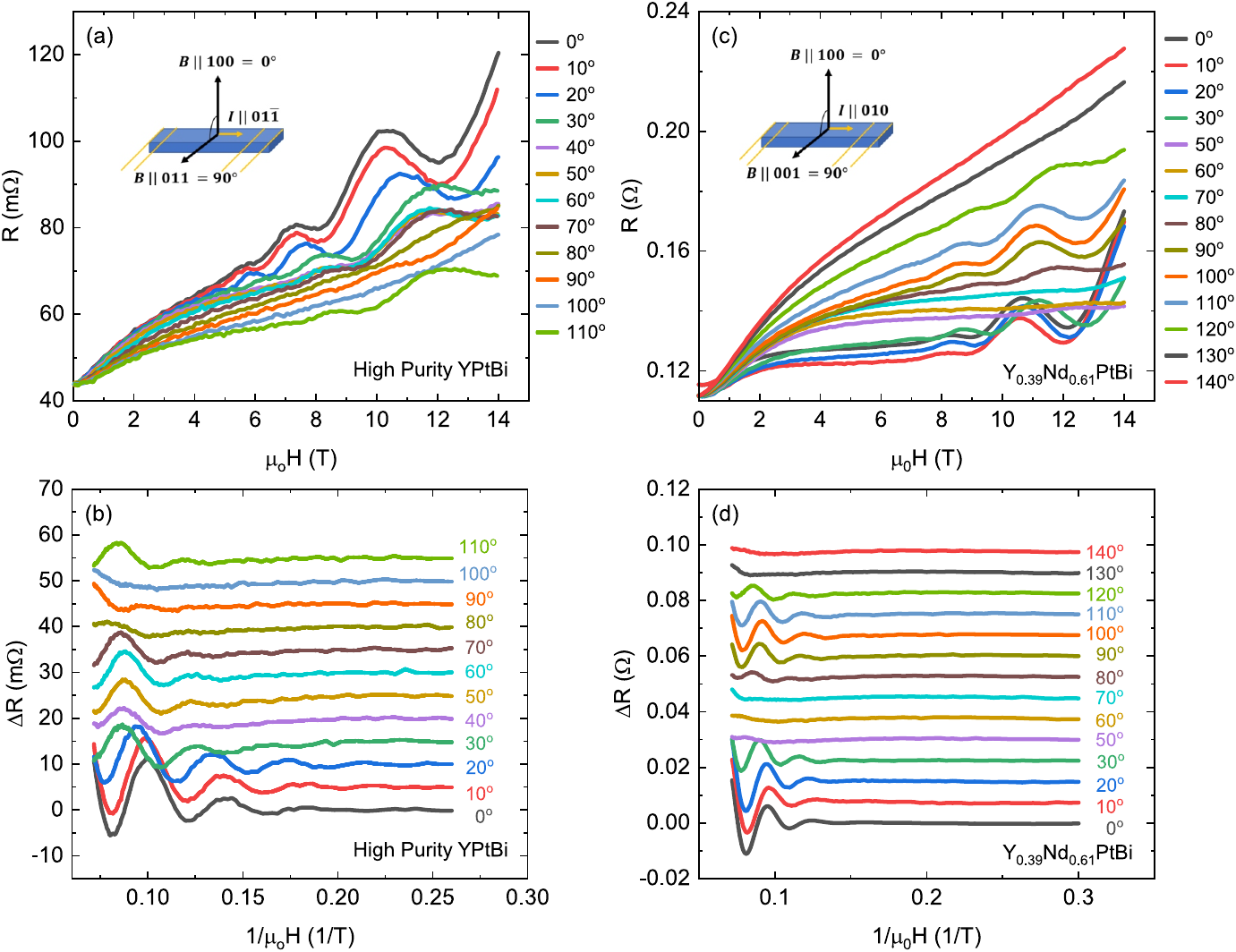}
    \caption{
    {\bf Angle-dependent probe of $j$=3/2 Fermi surface in YPtBi.}
    Rotation measurements on the high purity and (Y,Nd)PtBi samples. (a,c) Raw data at various angles. Inset shows the rotation axis of the field along the principal axes. (b,d) Subtracted data offset to show the oscillations at various angles clearly showing a suppression of amplitude around to 011 axis for both samples.}
    \label{fig:QO-HP}
\end{figure*}

Fig.~\ref{fig:Transport}(a) presents representative temperature dependent resistivity measurements from 300K to 2K for all four catetories of samples. Normal- and high-purity YPtBi,(Y,Nd)PtBi and disordered YPtBi all exhibit non-metallic behavior characterized by increasing resistivity on cooling, consistent with prior reports~\cite{ButchPRB}. In contrast, Ca-doped samples show metallic behavior consistent with the expectation of an increased carrier density, which will be elaborated on below. 
To understand the degree of metallicity, we extracted the inverse residual resistivity ratio (1/RRR), defined as $1/RRR = \rho(2K)/\rho(300 K)$. Both the high purity, (Y,Nd)PtBi and disordered samples exhibit significantly larger values of 1/RRR compared to standard YPtBi, suggesting either increased scattering, lowered carrier density, or a combination of both. 
For instance, in disordered samples, the inclusion of Au during the growth process introduces vacancies and site disorder thought to increase scattering, whereas the high purity YPtBi material is thought to have the lowest carrier density due to reduced impurity doping effects. 
However, to quantify and segregate these effects, we have performed both Hall effect and quantum oscillations experiments to directly probe the carrier densities.
Fig.~\ref{fig:Transport}(b) shows the Hall resistivity $\rho_{xy}$ as function of magnetic field at 2 K for all sample types. Across all samples $\rho_{xy}(H)$ is approximately linear as expected for the near-spherical single-band structure of YPtBi \cite{HKimQO-PRS}, allowing us to extract the Hall coefficient $R_H = R_{xy}t/\mu_0H$ from the slope, where $t$ is the sample thickness. Using the relation $R_H = 1/ne$, we estimate the carrier density $n$. Standard, high purity and disordered samples show only minor variations in slope, suggesting slight changes in carrier density. However, the Ca doped samples exhibit a significantly smaller slope corresponding to a substantial enhancement in carrier concentration. This result confirms that Ca substitution for Y acts as a strong hole donor, as expected by its divalent oxidation state.

To examine sample variability, we measured multiple crystals of each type and compiled statistics on transport parameters. Fig.~\ref{fig:Transport}(c) summarizes the residual resistivity $\rho_0 = \rho (2K)$ and 1/RRR of the samples. Both the disordered and high purity samples consistently show a higher $\rho_0$ and 1/RRR compared to standard YPtBi. This confirms that the inclusion of Au in the synthesis process indeed introduces disorder into the lattice, enhancing low temperature scattering, while the elevated resistivity and 1/RRR observed in high-purity sample can be attributed to a lowered carrier density. Fig.~\ref{fig:Transport}(d) further illustrates this by plotting carrier density as a function of 1/RRR. The disordered and (Y,Nd)PtBi samples have a similar carrier density ($n_{\text{disordered}} \approx 1.84 \times 10^{18}\,\text{cm}^{-3}$ and $n_{\text{(Y,Nd)PtBi}} \approx 2.63 \times 10^{18}\,\text{cm}^{-3}$ ) to YPtBi ($n_{\text{YPtBi}} \approx 2 \times 10^{18}\,\text{cm}^{-3}$~\cite{ButchPRB}), as confirmed by both Hall and quantum oscillation measurements. In contrast, high purity samples show a significantly reduced carrier density ($n_{\text{high purity}} \approx 0.7 \times 10^{18}\,\text{cm}^{-3}$), indicating that the purer raw materials suppress extrinsic doping and bring the Fermi level closer to the band touching point. Ca-doped samples, however, display a wide range of higher carrier concentrations ranging from $n \approx 16 \times 10^{18}$ to $n \approx 320 \times 10^{18}\,\text{cm}^{-3}$, up to two orders of magnitude larger than in YPtBi. These findings confirm that Ca doping increases the number of charge carriers.

Quantum oscillations are a direct, sensitive probe of the Fermi surface and the effective mass of the charge carriers. In standard YPtBi, clear quantum oscillations have been observed at fields below 14 T~\cite{HKimSciAdv}. However, increased impurity scattering in the disordered and Ca-doped samples leads to Dingle damping of the oscillation amplitude~\cite{Shoenberg}, making oscillations undetectable up to the same field range. To overcome this limitation we conducted high-field magnetoresistance measurements up to 41.5~T. 
Fig.~\ref{fig:QO}(a-d) display the raw magnetoresistance for the high purity, disordered, (Y,Ca)PtBi and (Y,Nd)PtBi samples respectively. 
Clear Shubnikov de Haas (SdH) oscillations are visible in the high purity samples, with oscillations persisting up to 40 K. The fast Fourier transform (FFT) of this data, shown in Fig.~\ref{fig:QO}(e), reveals a dominant frequency of $F \approx 22 \,\text{T}$, significantly lower than that of standard YPtBi ($F \approx 51 \,\text{T}$). The temperature dependence of the FFT amplitude shown in Fig.~\ref{fig:QO}(f) fits well to the Lifshitz-Kosevich (LK) expression~\cite{Shoenberg} yielding an effective mass of $m^* = 0.069m_e$ which is notably lighter than that of YPtBi ($m^* = 0.11m_e$). These results are consistent with a smaller Fermi surface and a lower Fermi energy.

Fig.~\ref{fig:QO}(a) presents the weaker quantum oscillations observed in disordered samples. Although the oscillations are less prominent due to increased scattering, the background-subtracted oscillatory component is indeed periodic in $1/\mu_\circ H$, as shown in the inset. The FFT in Fig.~\ref{fig:QO}(e) reveals a peak at $F \approx 54\, \text{T}$, similar to that of YPtBi. The extracted effective mass is $m^* = 0.11m_e$ matching that of YPtBi. These observations suggest that while disorder suppresses the quantum oscillation amplitude by enhancing scattering, it does not significantly shift the Fermi level or alter the carrier density, consistent with the Hall data described above.

Ca substitution, expected to increase hole carrier density, indeed induces a stark shift in the oscillation frequency. FFT analysis (Fig.~\ref{fig:QO}(f)) reveals a threefold increase in frequency up to $\sim 157\, \text{T}$, reflecting a substantial increase in the Fermi surface cross sectional area due to elevated carrier density. The LK fit yields an effective mass of $m^* = 0.18m_e$, indicating a heavier band structure at higher energies, likely associate with doping induced band filling. These results confirm that Ca doping raises the Fermi level significantly, modifying both the carrier density and the effective mass.
Quantum oscillations of (Y,Nd)PtBi crystals, shown in Fig.~\ref{fig:QO}(d), are also evident upon background subtraction, and exhibit a reduced frequency of $ \sim 33\, \text{T}$ again consistent with reduced carrier density derived from Hall effect measurements. The measured effective mass $m^* = 0.11m_e$ is  similar to that of standard YPtBi.

Finally, the effect of these perturbations on the high angular momentum
quasiparticle Fermi surface of YPtBi can be probed through unique signatures in Fermiology \cite{HKimQO-PRS}. First, a splitting of the primary oscillation frequency into two components (39~T and 50~T) was understood as a signature of inner and outer orbits of spin-split Fermi surfaces due to the absence of parity symmetry ~\cite{HKimSciAdv}. In our experiments, this effect has not been resolved, likely due to the limited field window in samples with larger Dingle factor, and the lower number of oscillation periods in the high purity case limiting our resolution. 
Second, a notable and readily observed consequence of the $j$=3/2 Fermi surface is a strong amplitude variation of oscillations in this otherwise isotropic system \cite{HKimQO-PRS}. 
As illustrated in Fig.~\ref{fig:QO-HP}, pronounced SdH oscillation amplitudes in both high purity YPtBi and (Y,Nd)PtBi show the expected strong angular variation as a function of transverse field rotation, with largest amplitude found for [100] orientation and strong suppression upon approach to the equivalent [110] direction in each case. FFT analysis confirms the frequency remains essentially constant through the amplitude change, consistent with prior work confirming the $j$=3/2 Fermi surface.


\begin{figure*}[]
    \centering
    \includegraphics[width=0.98\linewidth]{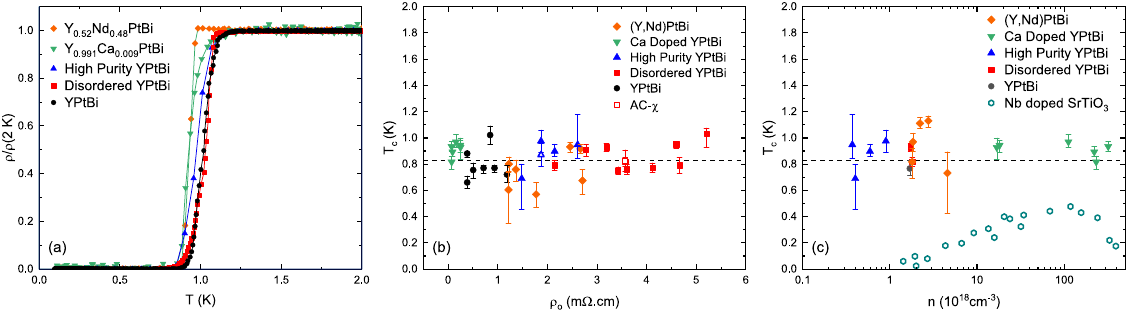}
    \caption{
    {\bf Insensitivity of $j$=3/2 superconductivity to disorder and carrier density tuning.} 
    (a) Normalized resistivity curves for the samples clearly showing no variation in transition temperature. (b) Various samples measured in resistivity showing no change in $T_c$ due to disorder. (c) $T_c$ vs carrier density showing no change in the transition temperature. }
    \label{fig:SC}
\end{figure*}

With representative characterizations of disorder, purity, substitution and doping effects on these samples, we now explore the impact on superconductivity in YPtBi.
Across all sample types studied in this work, superconductivity remains exceptionally tolerant. Fig.~\ref{fig:SC} summarizes the robust nature of the superconducting transition to these perturbations, including little change in the resistive $T_c$, whether plotted as a function of sample type (Fig.~\ref{fig:SC}(a)), residual resistivity (Fig.~\ref{fig:SC}(b)) or carrier density changes (Fig.~\ref{fig:SC}(c)).
Through the combination of ultra-high purity, induced disorder and hole-doping, this insensitivity spans a remarkable range of parameter space, from the lowest carrier densities ($\sim$10$^{17}$ cm$^{-3}$) to the highest scattering rates ($\sim$5~m$\Omega$cm). 

This places strong constraints on the mechanism responsible for superconducting pairing in this material, in particular in the cases of nonmagnetic disorder and carrier doping.
In conventional BCS superconductors with isotropic s-wave pairing, nonmagnetic disorder does not suppress superconductivity as a result of Anderson's theorem. 
However, in unconventional superconductors with a sign-changing gap structure, both magnetic and non-magnetic disorder act as strong pair-breakers, leading to rapid suppression of $T_c$ characterized by the Abrikosov-Gor'kov (AG) model~\cite{AGTheory,Larkin} 
The fact that superconductivity in YPtBi persists even in heavily disordered samples, where residual resistivity is strongly enhanced, suggests that the pairing state is either fully gapped and protected by symmetry, or rooted in an unconventional but disorder-resilient channel. One possibility is the presence of higher-angular momentum multi-component order parameters such as quintet (J = 2) or septet (J = 3) pairing states, which can exhibit complex gap structures and in some cases robustness to certain types of disorder~\cite{JYuJAP,cavanaghPRB}.

The lack of sensitivity to carrier density is a more profound observation, challenging the applicability of a conventional electron-phonon mediated pairing mechanism which typically exhibits a strong dependence on the density of states and is sensitive to changes in band filling.
For a conventional BCS superconductor, $T_c$ is expected to increase with increasing carrier density \cite{tinkham2004introduction}. While a few counter examples exist, such as unconventional dilute superconductors like Li-intercalated layered nitrides \cite{TaguchiLixZrNCl, Nakagawa-GateControlled},
these are cases where $T_c$ nevertheless shows a strong response to doping.
However, the lack of $T_c$ change over three orders of magnitude carrier density tuning (cf. Fig.~\ref{fig:SC}(c)) is unprecedented. Even in the case of 
the dilute superconductor SrTiO$_3$,
which has emerged as a canonical example of superconductivity at extremely low carrier densities \cite{BEC-BCS-Theory,Ealges1969},
there is a strong variation with doping that is evident in the comparison in Fig.~\ref{fig:SC}(c)).
The presence of superconductivity in both the dilute and heavily doped metallic regimes suggests that the pairing mechanism is not only robust but fundamentally distinct from any other known superconductor.

This behavior supports the idea that superconductivity in YPtBi is not tied to a fine-tuned Fermi surface instability or a high density of states at the Fermi level, but rather an alternative mechanism.
In conventional pairing, $T_c$ is primarily governed by the pairing amplitude and scales with the density of states at the Fermi level. In contrast, in the dilute limit where the Fermi energy becomes comparable to the pairing scale, superconductivity can enter a crossover regime in which fermion pairs form at temperatures well above the phase ordering temperatures and long range superconducting coherence is instead governed by superfluid phase stiffness $D_S \propto n_s/m^*$, where $n_s$ is superfluid density and $m^*$ is the effective mass.
This scenario has been heavily studied in the cuprate superconductors \cite{Emery1995}, where the parallel increase of $T_c$ and superfluid density in the underdoped regime is consistent with a phase ordering limited transition, while becoming insensitive to further increases in stiffness on the overdoped side \cite{Uemura-1989}. In particular, once the phase stiffness exceeds the minimum threshold required to establish global coherence, further increases in the $n_s/m^*$ do not lead to enhancement of $T_c$, indication a crossover from a stiffness limited to a pairing limited regime \cite{KIVELSON200261}. 
While YPtBi is inherently distinct from the cuprates, this phenomenology provides an important precedent for interpreting a saturation of $T_c$ in the presence of large changes in carrier density and Fermi surface volume.

Several aspects of YPtBi naturally place it near the crossover regime. The carrier densities of the undoped and high purity samples imply an extremely small Fermi energy, comparable to or moderately larger than the superconducting gap scale inferred from $T_c$. In this limit, coherence length is expected to be short and the size of the Cooper pairs can approach interparticle spacing, blurring the distinction between weakly bound Cooper pairs and tightly bound composite bosons. Similar physics has been extensively discussed in dilute superconductors and semimetals, most notably in SrTiO$_3$, which cannot be adequately described within weak coupling BCS theory and instead requires consideration of strong coupling, pairing fluctuations and the separation between pair formation and phase coherence scales \cite{SrTiO3-Review}. In this system, superconductivity persists down to carrier densities as low as 10$^{17}$ cm$^{-3}$, and the small Fermi energy places the system close to a BCS-BEC crossover regime in which phase coherence plays a central role in determining $T_c$ \cite{BEC-BCS_Review}. Although the microscopic pairing mechanism in SrTiO$_3$ differs from that in YPtBi, the underlying energetic hierarchy, small $E_F$, robust pairing and potentially phase stiffness-limited superconductivity, provides a useful and well established point of comparison.

Within this picture, $T_c$ is not limited by pair formation but by the ability of the system to establish phase coherence. Notably, phase stiffness can be weakly sensitive to non-magnetic disorder and only weakly dependent on the details of the Fermi surface, provided sufficient density of paired carriers exists. This provides a natural explanation for the striking insensitivity of $T_c$ in YPtBi to disorder, carrier concentration and even substantial changes to the Fermi surface volume by Ca doping.

Importantly, the phase stiffness perspective is not mutually exclusive with unconventional high spin pairing. On the contrary, pairing between $j$=3/2 quasiparticles can naturally yield strong pairing amplitudes even at low carrier densities, while the resultant multicomponent order parameter might be relatively insensitive to impurity scattering. In this sense, YPtBi may represent a rare example where high angular momentum, spin-orbit-entangled pairing coexists with crossover physics typically associated with dilute superconductors, placing it in a regime where superconductivity is stabilized by robust pair formation but limited primarily by phase stiffness.



These findings establish YPtBi as a highly unusual superconductor where superconductivity is a remarkably robust ground state. The interplay between low carrier density, strong spin-orbit coupling, band inversion and potentially phase stiffness limited coherence create a fertile ground for unconventional pairing. This work not only deepens our understanding of the complex relationship between electronic structure and superconductivity in YPtBi, but also provides a foundation for future theoretical and experimental studies aimed at uncovering the topological nature of its superconducting state.

\section{Methods}

YPtBi single crystals were grown using the molten Bi flux method. Y, Pt, and Bi in a molar ratio of 1:1:20 were placed in an alumina crucible and sealed in a quartz tube under an argon atmosphere. The excess Bi acts as a self-flux. The mixture was heated to 1100\,$^{\circ}$C at a rate of 50\,$^{\circ}$C/hour, held at that temperature for 12 hours to ensure homogenization, and then slow-cooled to 520\,$^{\circ}$C at 3\,$^{\circ}$C/hour before centrifuging to separate the crystals from the flux. Unless otherwise noted, measurements on these as-grown samples are represented by black circles or lines throughout this work.

High-purity crystals were grown using higher-purity starting materials: 99.99\% Y, 99.999\% Pt, and 99.9999\% Bi at the Max Planck Institute for Chemical Physics of Solids. Standard-purity samples used 99.9\% Y and Pt and 99.99\% Bi. Measurements on these high-purity samples are represented by blue triangles or lines.

To introduce additional disorder without altering chemical composition, Au powder was added to the growth in varying quantities, up to a maximum of 20\%. Elemental analysis via energy-dispersive spectroscopy (EDS) and wavelength-dispersive spectroscopy (WDS) showed no detectable Au in the crystals. Because these techniques are surface-sensitive, we also performed inductively coupled plasma (ICP) analysis, which dissolves the entire crystal to probe its bulk composition. ICP confirmed the absence of Au, indicating that it is not chemically incorporated into the lattice. This suggests that Au acts solely as a source of disorder, introducing additional scattering centers without affecting the chemical potential. This conclusion is further supported by our electrical transport measurements. These disordered samples are indicated by red squares or lines in the figures.

Ca-doped samples were prepared by introducing Ca into the melt, in concentrations up to 50\% relative to the Y molar amount. X-ray fluorescence (XRF) and EDS confirm the presence of Ca in the crystals, though the incorporated concentration is significantly lower than the nominal, reaching a maximum of $\sim$6\%. Despite this, clear signatures of hole doping are observed in measurements of resistivity, Hall effect, and quantum oscillations. These Ca-doped samples are denoted by green triangles or lines throughout this work.

Nd-substituted samples were prepared by introducing Nd into the melt, in concentrations up to 50\% relative to the Y molar amount. The concentration of Nd was confirmed by magnetic susceptibility measurements. These Nd substituted samples are denoted by green triangles or lines throughout this work.

Powder x-ray diffraction (XRD) measurements on the samples confirm that they retain the YPtBi crystal structure, with diffraction peaks matching published data. 

Electrical resistivity samples were prepared by polishing crystals into bar-shaped samples and attaching gold wire leads with silver paste, and subsequently measuring in a Quantum Design Physical Property Measurement System (PPMS) equipped with a 14T magnet for resistance and magnetoresistance measurements. To probe the superconducting state we used the adiabatic demagnetization refrigerator (ADR) insert with the PPMS, applying a field procedure (ramping to 2.5 T then zero) to cool samples to  100 mK by adiabatic demagnetization refrigeration. We measure the resistivity as the system warms up from 100 mK at zero field. High magnetic field measurements were done with a 41.5 T resistive magnet at the National High Magnetic Field Laboratory (NHMFL).

To investigate the strong anisotropy of oscillation amplitude due to the $j$=3/2 electronic states in YPtBi, we carried out comprehensive angle dependent quantum oscillation measurements in the high purity sample and Nd-substituted samples. Samples were first aligned with Laue x-ray diffraction to ensure precise crystallographic alignment. Following alignment, resistivity measurements were performed at 2 K under magnetic fields up to 14 T. The electrical current was applied along [011] crystallographic direction for the high purity sample and [010] axis for the Nd substituted samples. To systematically study the angular dependence, the magnetic field was rotated in discrete 10$^\circ$ increments. Throughout the rotation, the magnetic field was kept perpendicular to the current.

\begin{acknowledgments}
Research at the University of Maryland was supported by
the Gordon and Betty Moore Foundation’s EPiQS Initiative Grant No. GBMF9071 (materials synthesis),
the U.S. National Science Foundation Grant No. DMR2303090 (sample preparation), 
the Department of Energy, Office of Basic Energy Sciences Award No. DE-SC-0019154 (experimental measurements), 
the NIST Center for Neutron Research, and the Maryland Quantum Materials Center. The research work at Max Planck Institute for Chemical Physics of Solids was financially supported by the Deutsche Forschungsgemeinschaft (DFG, German Research Foundation) through SFB 1143 (project ID 24731007), QUAST (project ID FOR 5249), the  Würzburg-Dresden Cluster of Excellence ctd.qmat – Complexity, Topology and Dynamics in Quantum Matter (EXC 2147, project-id 390858490), and the European Union through EXQIRAL (No. 101131579).

\end{acknowledgments}

\bibliography{YPtBi-RobustSC}

\end{document}